\documentclass[twocolumn,floatfix,prb,aps,showpacs]{revtex4-1}
\usepackage{graphicx,amsmath,amssymb}

\begin{document}

\newcommand{\bdelta}{\mbox{\boldmath$\delta_{12}$}}
\newcommand{\bdeltaast}{\mbox{\boldmath$\delta^\ast_{12}$}}

\title{Charge-density-wave states in double-layer graphene structures in a high magnetic field}

\author{Csaba T\H oke$^1$}
\author{Vladimir I. Fal'ko$^2$}
\affiliation{$^{1}$BME-MTA Exotic Quantum Phases ``Lend\"ulet" Research Group,
Budapest University of Technology and Economics,
Institute of Physics, Budafoki \'ut 8, H-1111 Budapest, Hungary}
\affiliation{$^{2}$Department of Physics, Lancaster University, Lancaster LA1 4YB, United Kingdom}
\date{\today}

\begin{abstract}
We study the phases of correlated charge-density waves that form in a high magnetic field in two parallel
graphene flakes separated by a thin insulator.
The predicted phases include the square and hexagonal charge-density-wave bubbles,
and a quasi-one-dimensional stripe phase.
We find that the transition temperature for such phases is within the experimentally accessible range
and that formation of interlayer-correlated states produces a negative compressibility contribution
to the differential capacitance of this system.
\end{abstract}


\maketitle

\section{Introduction}

Interaction-coupled parallel two-dimensional electron gases (2DEGs) in semiconductor structures 
are interesting objects from the point of view of electron-electron correlation effects: 
interlayer drag,\cite{coulombdrag} excitonic superfluidity,\cite{exfluid} and even-denominator fractional quantum Hall 
states.\cite{newfqhe}  The creation of van der Waals--coupled 
graphene--hexagonal boron nitride--graphene (G/hBN/G) multilayers, 
by mechanical exfoliation and transfer,\cite{Ponomarenko,Amet}  
offers a system where the interlayer correlations develop at 
elevated temperatures and in earlier inaccessible parametric regimes
because of the extreme thinness of both the conducting layers and the barrier.

\begin{figure}[htbp]
\begin{center}
\includegraphics[width=\columnwidth,keepaspectratio]{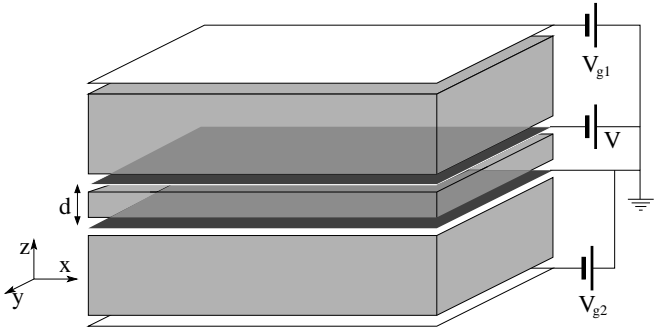}
\end{center}
\caption{\label{device}
Sketch of the device measurement scheme.
}
\end{figure}

Motivated by the emergence of this experimentally accessible system, we model the
charge-density waves (CDWs) in two independently contacted\cite{tunneling,Ponomarenko,Amet} parallel
graphene layers (1 and 2) with carrier densities $\rho_{1,2}$ corresponding to almost integer filling
of the $|n|=1,2,\dots$ orbital Landau levels (LLs) of electrons 
in a high perpendicular magnetic field $B$ (see Fig.~\ref{device}).
In a conventional 2DEG in high-quality
semiconductors,\cite{Fukuyama,Fogler,Moessner,Haldane,stripesdetail,liquidcrystal,reviews} as well as in a
single graphene layer,\cite{Joglekar} electrons in partially filled higher Landau levels
have been predicted to form a CDW state: a bubble or stripe phase. 
The formation of such symmetry-broken states by electrons in higher LLs is promoted by the 
spatial structure of LL wave functions, which for $|n|\ge1$ have minima in the electron density.

In this paper we show that similarly to the Wigner crystal in two-layer electron systems\cite{bilayerwc1,bilayerwc2,bilayerwc3}, 
the interlayer correlations of electrons in G/hBN/G heterostructures are able to produce a multiplicity of CDW states 
in the experimentally feasible range of interlayer separtions $d$ comparable to
the electron cyclotron radius $R^{(n)}_c \sim \sqrt{2|n|+\delta_{n0}}\ell$ ($\ell=\sqrt{\hbar/eB}$). 
Similarly to the earlier theories of the Wigner crystal in two-layer electron systems,\cite{bilayerwc1,bilayerwc2,bilayerwc3} we compare the numerically calculated 
ground-state energies of CDW states with the 
high-symmetry Bravais lattices---rhombic, hexagonal, rectangular, and square---taking
into account the layer-dependent charge-density structures 
within the CDW supercell, and find the most favorable phase.

The paper is structured as follows.
In Sec.~\ref{results}, we present our main result, i.e.,
the phase diagram of CDW states in a graphene double-layer system.
In Sec.~\ref{methods}, we explain the methods and approximations utilized to obtain this phase diagram.
In Sec.~\ref{discussion}, we discuss in detail its features, spell out the experimental connections, and
provide quantitative predictions related to the electronic compressibility and estimated critical temperatures.
Section \ref{summary} summarizes our main points.

\section{Results}
\label{results}

\begin{figure}[htb]
\includegraphics[width=0.85\columnwidth,keepaspectratio]{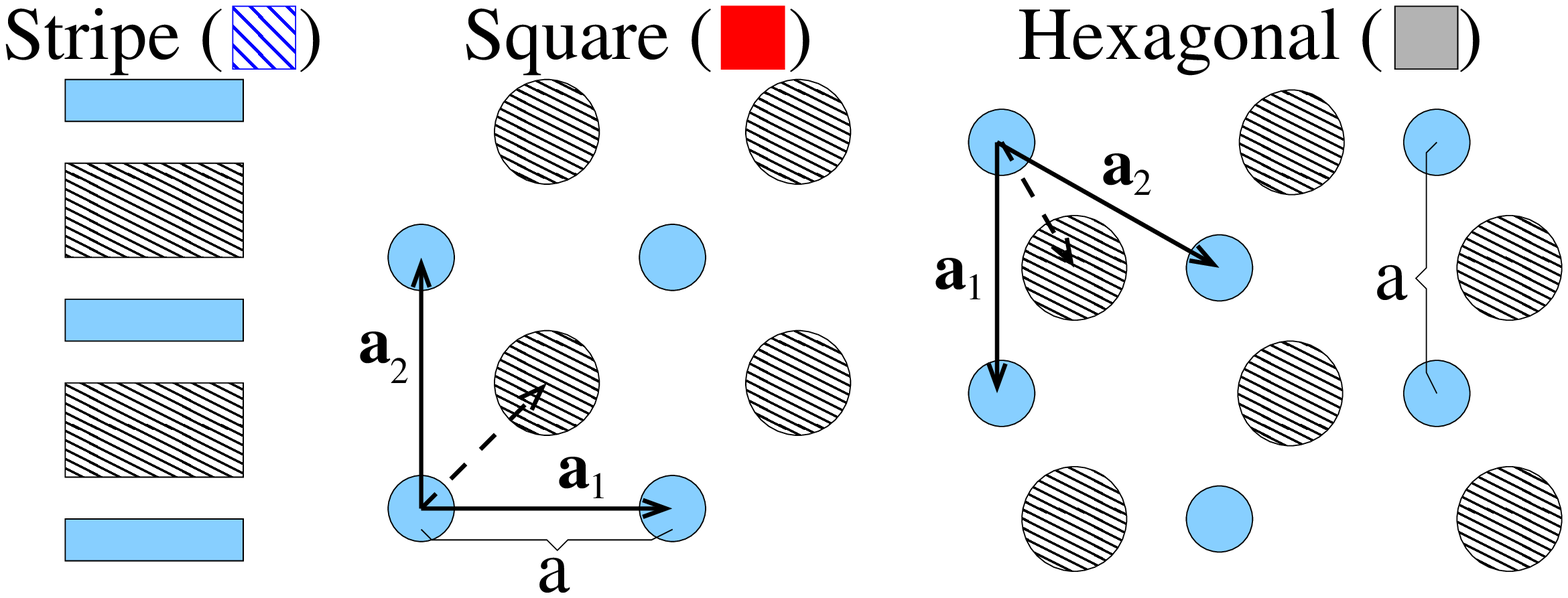}
\begin{flushleft}
\includegraphics[width=0.75\columnwidth]{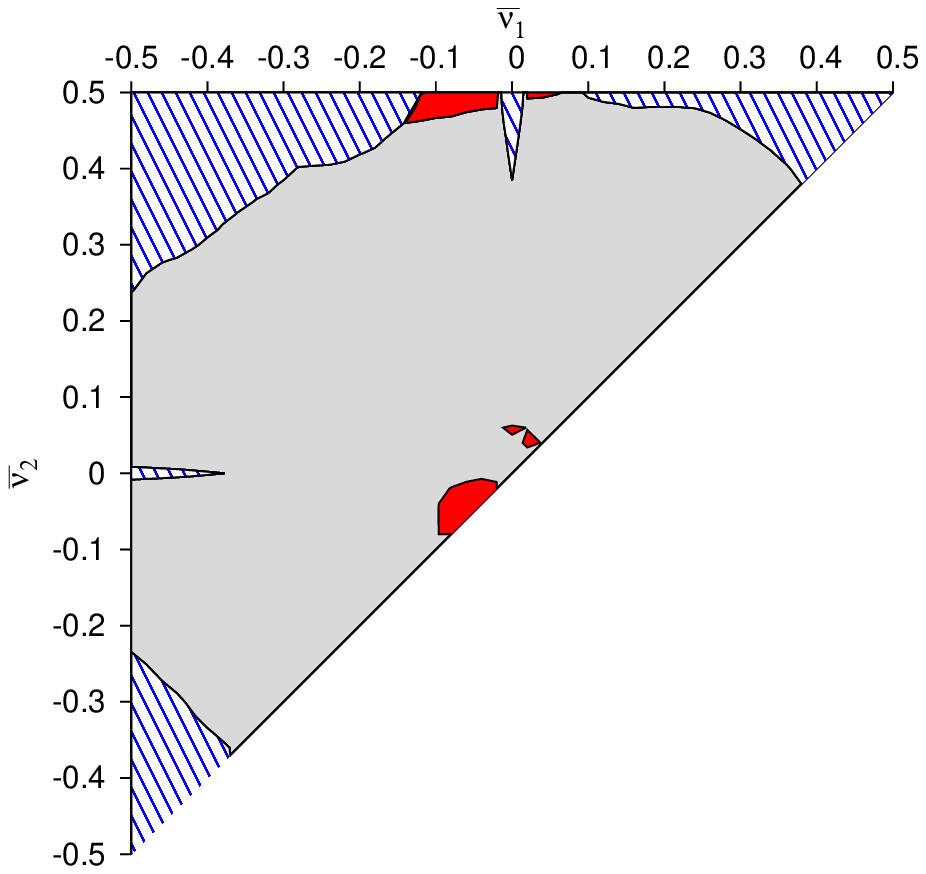}%
  \setlength{\unitlength}{0.01\columnwidth}
  \begin{picture}(100,20)    
    \put(10,10.5){\includegraphics[width=80.5\unitlength]{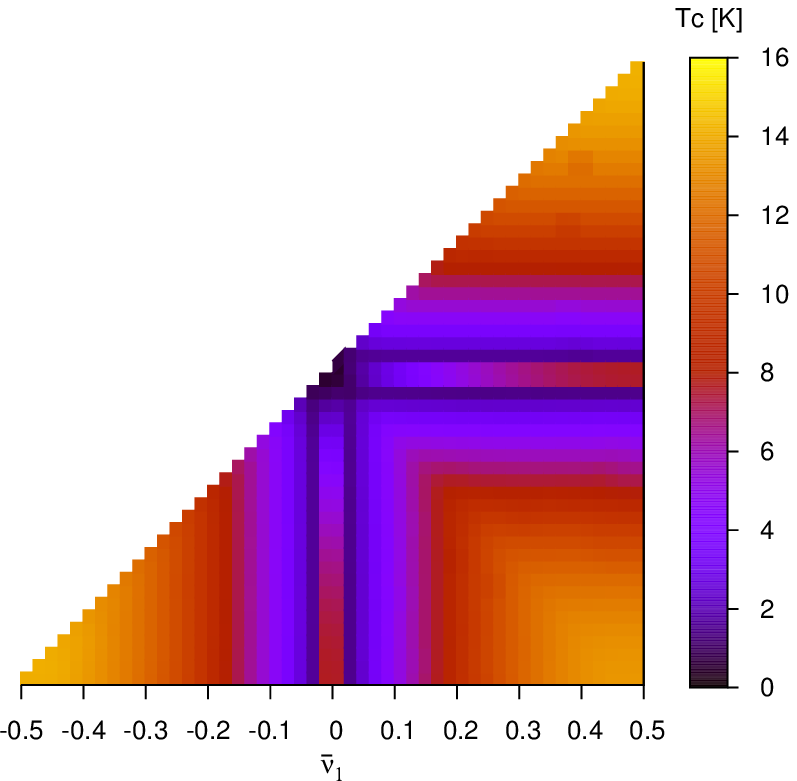}}
  \end{picture}
\end{flushleft}
\caption{\label{phases} (Color online) 
CDW phases in a graphene structure with 4 nm spacer (equivalent to 12 hBN layers) around $\nu_{1,2}\approx 6$ at $B=10$T.  
The left half panel shows the appearance of a stripe phase, square, and hexagonal CDW as a function of $\overline\nu_i=\nu_i-6$;  
the right half panel shows the ordering transition temperature $T_c$ (in Kelvin) determined using a mean-field-theory approach. }
\end{figure}

Figure \ref{phases} presents the phase diagram of two parallel graphene sheets with a small partial filling factor, 
$\overline\nu_i=\nu_i-6$ ($\nu_i=2\pi\ell^2\rho_i$, $|\overline\nu_i|\le0.5$), of electrons in $n=2$ 
or holes in $n=1$ LLs in each layer, found using a $T=0$ mean-field approach. This phase diagram
includes islands of stability of the stripe phase, predicted earlier for monolayer graphene,\cite{Joglekar} CDW states of 
electron-hole pairs with a square lattice, and hexagonal and square two-layer CDWs. 
In addition, we identify parametric intervals where  
electron-electron correlations generate a negative contribution to the overall compressibility of the two-layer system and 
a quantum correction to its classical geometry-defined electrical capacitance (Fig.~\ref{comp}).
In this analysis, we take into account two important features of the two-layer graphene system: 
(i) electrons in higher Landau levels 
in graphene have different envelope functions on the two sublattices of the honeycomb lattice,
which affects the form factor  of the electron-electron (e-e) interaction,
and (ii) e-e interaction is also screened by the Fermi sea of the $\pi$-band electrons 
in the valence and conduction bands in both layers, which reduces a naively expected 
enhancement of the e-e correlation effects in the two-layer graphene system based on a closer spacing of graphene flakes 
and a smaller dielectric constant of the surrounding medium than is possible in semiconductor heterostructures.

\section{Methods}
\label{methods}

\subsection{Screening of the electron-electron interaction}

The polarization of electron Fermi sea in the valence and conduction bands of graphene screens the e-e interaction,
converting\cite{Mele} the bare intra- and interlayer interactions,\cite{footnoteE}
\begin{gather}
V_{11}(\mathbf q)=V_{22}(\mathbf q)=2E_0\pi/q, \qquad  
E_0=e^2/(4\pi\epsilon_0 \sqrt{\epsilon_\perp\epsilon_\parallel}\ell), \nonumber\\
V_{12}(\mathbf q)=V_{21}(\mathbf q)=(2E_0\pi/q)e^{-dq\sqrt{\epsilon_\parallel/\epsilon_\perp}},
\end{gather}
into the random-phase-approximation (RPA) expression
\begin{gather}
\mathbf{\widetilde V}(\mathbf q) = \left[1-\mathbf{V}(\mathbf q) 
\begin{pmatrix}
\Pi_{1} & 0 \\ 0 & \Pi_{2}
\end{pmatrix}
\right]^{-1}\mathbf{V}(\mathbf q),\nonumber\\
\mathbf V(\mathbf q)=\begin{pmatrix}
V_{11}(\mathbf q) & V_{12}(\mathbf q) \\
V_{21}(\mathbf q) & V_{22}(\mathbf q)
\end{pmatrix},
\end{gather}
where the static polarization $\Pi_i \equiv \Pi_i(\mathbf q,\omega=0)$ of Dirac electrons in the $i$th
graphene layer is defined as
\begin{equation}
\Pi_i = \frac{1}{\sqrt{2}\pi\ell v_F}
\sum_{\sigma\tau}
\sum_{n>n_{i\sigma\tau}}
\sum_{\tilde n\le n_{i\sigma\tau}}
\frac{|\tilde F_{n}^{\tilde{n}}(\mathbf q)|^2}{\text{sgn}(n)\sqrt{|n|} - \text{sgn}(\tilde{n})\sqrt{|\tilde{n}|}},
\nonumber
\end{equation}
with $n_{i\sigma\tau}$ standing for the index of the highest completely filled LL of spin $\sigma$ and valley $\tau$ in layer $i$, $v \approx 10^6$ m/s, and
\begin{multline}
\tilde F_{n}^{\tilde n}(\mathbf q) = \delta_{n0}\delta_{\tilde{n}0}F_{n}^{\tilde{n}}(\mathbf q)
+ \frac{\delta_{n0}+\delta_{\tilde{n}0}-2\delta_{n0}\delta_{\tilde{n}0}}{\sqrt 2}F_{|n|}^{|\tilde{n}|}(\mathbf q) +\\
+ \frac{(1-\delta_{n0})(1-\delta_{\tilde{n}0})}{2}\left(F_{|n|}^{|\tilde{n}|}(\mathbf q)+\text{sgn}(n\tilde{n})F_{|n|-1}^{|\tilde{n}|-1}(\mathbf q)\right); \\
F_{\tilde{n} \ge n}^{n}(\mathbf q)=\sqrt\frac{n!}{\tilde{n}!}\left(\ell\frac{q_y-iq_x}{\sqrt2}\right)^{\tilde{n}-n}L_n^{\tilde{n}-n}\left(\frac{q^2\ell^2}{2}\right)
e^{-\frac{q^2\ell^2}{4}}; \\
F_{\tilde{n}<n}^{n}(\mathbf q)=[F_{n}^{\tilde{n}}(-\mathbf q)]^\ast.
\end{multline}
Here, $L^m_n(z)$ are the associated Laguerre polynomials.
We will calculate $\Pi_i(\mathbf q,\omega=0)$ using a LL cutoff $n_{\text{max}}=5400\text{ T}/B$,
corresponding to the band-width in graphene.
For $q\ell >1$, the RPA results for $\mathbf{\widetilde V}(q)$ with or without a magnetic field\cite{Wunsch} hardly differ, whereas for $q\ell<1$,
the dominant contribution to $\Pi(\mathbf q,0)$ comes from the dipolar matrix elements between LLs.\cite{AG}

\subsection{The cohesive energy}

The cohesive energy of the CDW phases, defined as the difference between the energy of the CDW state and
the energy of the uniform electron liquid,\cite{Fogler} is calculated in the mean-field approximation.
Here we assume that the exchange interaction of electrons in a lightly filled LL spontaneously breaks the spin and valley
degeneracies, and the electrons (holes) in the partly filled LL are fully spin/valley polarized.
For fully polarized electrons, the charge-density waves are characterized\cite{Fukuyama,Fogler,Moessner} by
the guiding-center density,
\begin{gather}
\Delta_i(\mathbf q)=\left\langle \hat{\rho}_i (\mathbf q)\right\rangle, \\
\hat{\rho}_i (\mathbf q)\equiv \frac{1}{N_\phi}\sum_pe^{-iq_xp\ell^2} \hat a^\dag_{i,p+q_y/2}\hat a_{i,p-q_y/2},
\end{gather}
where $N_\phi$ is the number of flux quanta piercing the sample and $\hat a^\dag_{i,p}$ are creation operators of spin/valley polarized electrons (holes) in a state with $y$-momentum $p$ in a partially filled Landau level (here, we use the Landau gauge $\mathbf A=-Bx\mathbf{\hat y}$). Using the same ansatz for the Fourier harmonics of the CDW order parameter as in the earlier studies of stripe and bubble phases,\cite{Fukuyama,Fogler,Moessner,Haldane,stripesdetail,Joglekar} we describe the CDW order parameters as
\begin{gather}
\Delta^S_{i}(q_m) = e^{\pm \frac{i}{2}\bdelta\cdot\mathbf q_m}\frac{2}{a q_m}\sin\left(\frac{\overline\nu_i aq_m}{2}\right), \\
\Delta^B_{i}(\mathbf q _{m_1,m_2})=
e^{\pm \frac{i}{2}\bdelta\cdot\mathbf q  _{m_1,m_2}}\frac{2\pi R_{i}}{Aq_{m_1,m_2}}J_1(R_{i} q_{m_1,m_2}),\nonumber
\end{gather}
where, for the stripe phase (S) with period $a$, $q_m=2m\pi /a$, and for the bubble phases (B) with basis Bravais vectors $\mathbf{a_{1,2}}$, $\mathbf q _{m_1,m_2} = 2\pi (m_1\mathbf{a_{1}} + m_2\mathbf{a_{2}}) \times \mathbf{l_{z}}/A$.
Here, $A=|\mathbf a_1\times\mathbf a_2|$ is the area of unit cell of a 2D lattice CDW,
$R_i=\sqrt{\frac{A|\overline\nu_i|}{\pi}}$, $+$ and $-$ stand for layers $i=$1 and 2, respectively,
$\bdelta$ is the relative shift of the CDW sublattices in the two layers,
and $\Delta_i(-\mathbf q)=\Delta_i^\ast(\mathbf q)$.

Following Koulakov \textit{et al.},\cite{Fogler}
we evaluate the cohesive energy of a CDW state of electrons in the two parallel graphene flakes as
\begin{multline}
\label{coh}
E_{\text{coh}}=\frac{\ell^{-2}/4\pi}{|\overline\nu_1|+|\overline\nu_2|}
\sum_{m_1,m_2}\left[\sum_{i=1,2}|\Delta_i(\mathbf q _{m_1,m_2})|^2\tilde u_{HFi}(\mathbf q _{m_1,m_2})+\right.\\
+ \left. {\Delta_1}^{\ast}(\mathbf q _{m_1,m_2})\Delta_2(\mathbf q _{m_1,m_2})\tilde u^{\text{inter}}_H(\mathbf q _{m_1,m_2})\right],
\end{multline}
where $m_1^2+m_2^2 \ne 0$, and the Hartree-Fock, Hartree, and exchange potentials, respectively, are defined as  
\begin{gather}
\tilde u_{HFi}(\mathbf q)=\tilde u_{Hi}(\mathbf q) - 2\pi\ell^2 u_{Fi}(q\ell^2), \nonumber\\
\tilde u_{Hi}(\mathbf q)=\widetilde V_{ii}(q)|\tilde F^{n_i}_{n_i}(\mathbf q)|^2, \nonumber\\
u_{Fi}(\mathbf x)=\int\frac{d^2q}{(2\pi)^2}e^{i\mathbf q\cdot\mathbf x}\tilde u_{Hi}(\mathbf q), \nonumber\\
\tilde u^{\text{inter}}_H(\mathbf q)=\widetilde V_{12}(q)\tilde F_{n_1}^{n_1}(\mathbf q)\tilde F_{n_2}^{n_2}(\mathbf q).
\end{gather}

\section{Discussion}
\label{discussion}

\begin{figure}[htbp]
\begin{center}
\includegraphics[width=\columnwidth,keepaspectratio]{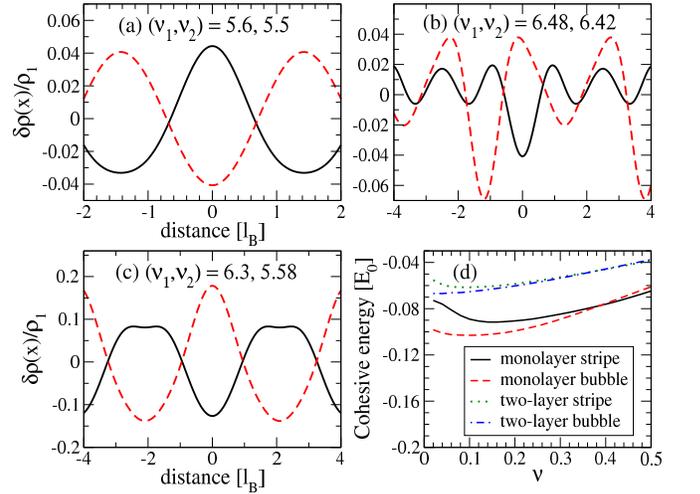}
\caption{\label{zoom}(Color online)
(a)--(c) The oscillatory part of the charge-density profile in units of $\rho_1=(2\pi\ell^2)^{-1}$, for the stripe phase.
(d) Comparison of the cohesive energies for monolayer and double layer CDWs for $\overline\nu_1=\overline\nu_2$. }
\end{center}
\end{figure}

\subsection{The phase diagram}

The phase diagram of the CDW states is found by minimizing (see Supplementary Material [\onlinecite{supp}])
its cohesive energy separately for the following:
stripes varying period $a$, CDW with rectangular Bravais lattice for which we vary the periods $a_1$ and $a_2$,
and a rhombic CDW for which we vary the angle between $\mathbf{a_{1}}$ and $\mathbf{a_{2}}$;
the square and hexagonal CDWs arise as special cases.
For each lattice, we also vary the mutual shift $\bdelta$ as an additional parameter to find the energy minimum. 
We find that, systematically, the lowest energy phases are: (a) quasi-one-dimensional CDW (stripe phase),
(b) a square lattice CDW, and (c) in the largest part of the phase diagram, the hexagonal phase.\cite{supp}
The resulting phase boundaries found for $T=0$ are shown in Fig.~\ref{phases} (lines where the energies of different CDW states coincide). 
The actual charge-density profile, which has additional structure as compared to the guiding center density due to the density profile of
Landau orbitals, is illustrated in Fig.~\ref{zoom}(a)--\ref{zoom}(c) for the stripe phase; similar behavior has been identified in the cases of square and hexagonal Bravais lattices.

The phase diagram in Fig.~\ref{phases} shows that in the range of high densities of added electrons or holes, the stripe phase, correlated between the two layers by the interlayer interaction, is preferable (in analogy to a single graphene layer\cite{Fogler,Joglekar}).
Moderate density imbalance does not destroy this phase.
If one layer is undoped, e.g., $\overline\nu_2=0$, then the stripe in the other is stable for $|\overline\nu_1|\gtrsim0.4$,
but any small $\overline\nu_2$ drives the systems to the hexagonal phase due to the great advantage of
Wigner-crystal-like ordering in layer 2.

There is a small interval of stability of square lattice CDW near $\nu_1=\nu_2=6$.
This structure arises because in two nearby layers the sparse packing of two
interlacing hexagonal lattices, i.e., a honeycomb lattice, means a disadvantage relative to the square, tipping
the delicate balance in favor of the latter.\cite{bilayerwc1,bilayerwc2}

In most of the phase space, the hexagonal CDW is promoted by the interlayer correlation effects in contrast
with single-layer graphene where the stripe CDW has been predicted.\cite{Joglekar}
However, as shown in Fig.~\ref{zoom}(d), the cohesive energies in the two-layer system are smaller in magnitude (less negative), which is due to the detrimental effect of the stronger screening of the Coulomb repulsion by the Fermi sea of electrons in remote Landau levels, which overcomes the stabilizing effect of the interlayer correlations.

\subsection{The critical temperature}

We estimate the critical temperature for each of these phases by a Landau-Ginzburg mean-field theory. 
For this, we calculate the free energy $\delta F$ of electrons in the CDW state,
substracting the energy of the uniform liquid,
\begin{multline}
\hat H^{\text{HF}}=\frac{1}{4\pi\ell^2}\sum_{\mathbf q = \mathbf q _{m_1,m_2}}\sum_{i=1,2}
\left[\tilde u_{HFi}(\mathbf q)
\frac{\hat\rho_i(\mathbf q)\Delta_i^\ast(\mathbf q)}{\tilde F_{n_in_i}(\mathbf q)}+\text{h.c.}+\right.\\ 
\left.\tilde u_{H}^{\text{inter}}(\mathbf q)
\frac{\hat\rho_i(\mathbf q)\Delta_{3-i}^\ast(\mathbf q)}{\tilde F_{n_in_i}(\mathbf q)}+\text{h.c.}-\right.\\
\left.-N_{\phi}|\Delta_i(\mathbf q)|^2-N_{\phi}\Delta_i(\mathbf q)\Delta^\ast_{3-i}(\mathbf q)
\right].
\end{multline}
Then, we analyze the temperature dependence of the free-energy difference term that is quadratic in the order parameter,
\begin{multline}
\delta F \approx -\frac{N_\phi}{2\pi}\left(U_{11} |\Delta_1|^2 + U_{22}|\Delta_2|^2 + 2U_{12}\Re \Delta_1^*\Delta_2\right), \\
U_{11}=\tilde u_{HF1}(\mathbf q) +
\frac{\gamma_1\tilde u^2_{HF1}(\mathbf q) +\gamma_2\left(\tilde u_{H}^{\text{inter}}(\mathbf q)\right)^2}{2\pi k_BT},\\
U_{22}=\tilde u_{HF1}(\mathbf q) +
\frac{\gamma_1\left(\tilde u_{H}^{\text{inter}}(\mathbf q)\right)^2 +\gamma_2\tilde u^2_{HF2}(\mathbf q)}{2\pi k_BT},\\
U_{12}=\tilde u_{H}^{\text{inter}}(\mathbf q)\left(1 +
\frac{\gamma_1\tilde u_{HF1}(\mathbf q) +\gamma_2\tilde u_{HF2}(\mathbf q)}{2\pi k_BT}\right),\\
\gamma_1=\overline\nu_1(1-\overline\nu_1),\quad\gamma_2=\overline\nu_2(1-\overline\nu_2). 
\end{multline}

For the stripe and square CDWs, where symmetry rules out any third-order invariants of the order parameter, the CDW transition is of the second order. Then, the above expression can be used to find the critical temperature $T_c$ of the phase transition: such temperature that $\delta F$ becomes negative at $T<T_c$.
Naturally, the instability is always due to the shortest few equivalent reciprocal lattice vectors.
For the stripe and square phases, the above method overestimates the critical temperature,
since melting would be dominated by the defects and, in ideally clean systems, the transition is of the Kosterlitz-Thouless type.
 
For the hexagonal CDW phase, symmetry allows for cubic terms in the Ginzburg-Landau theory.
In this case, we expect the phase transition to be a weak first-order transition,
so that the above procedure would underestimate its critical temperature:
the $T_c$ we get corresponds to the temperature of the absolute instability of the CDW state.
In the bottom half panel in Fig.~\ref{phases}, we plot the value of $T_c$ for $B=10$ T and
a barrier consisting of 12 hBN layers ($d/\ell=0.49$).
The calculated value of $T_c$ for a hexagonal CDW (Fig.~\ref{phases}) is, interestingly,
a nonmonotonic function of the filling factors.
The second interesting feature of the two-layer hexagonal CDW is
the twofold degeneracy related to the broken inversion symmetry: this degeneracy suggests the existence of two types of domains in the CDW ``crystal'' and domain boundaries, which can be pinned by disorder.

\begin{figure}[htbp]
\begin{center}
\includegraphics[width=0.49\columnwidth,keepaspectratio]{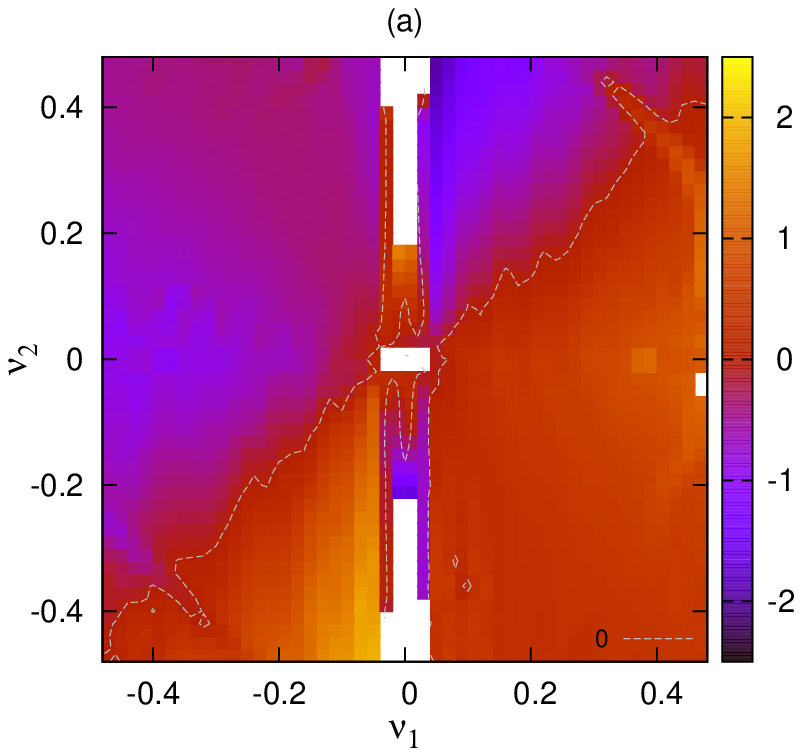}
\includegraphics[width=0.49\columnwidth,keepaspectratio]{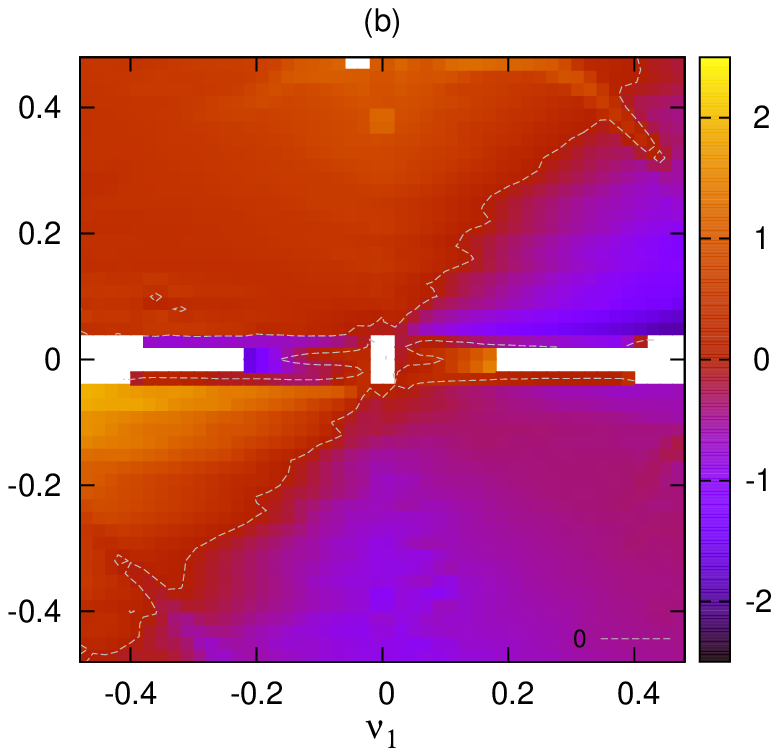}

\includegraphics[width=0.49\columnwidth,keepaspectratio]{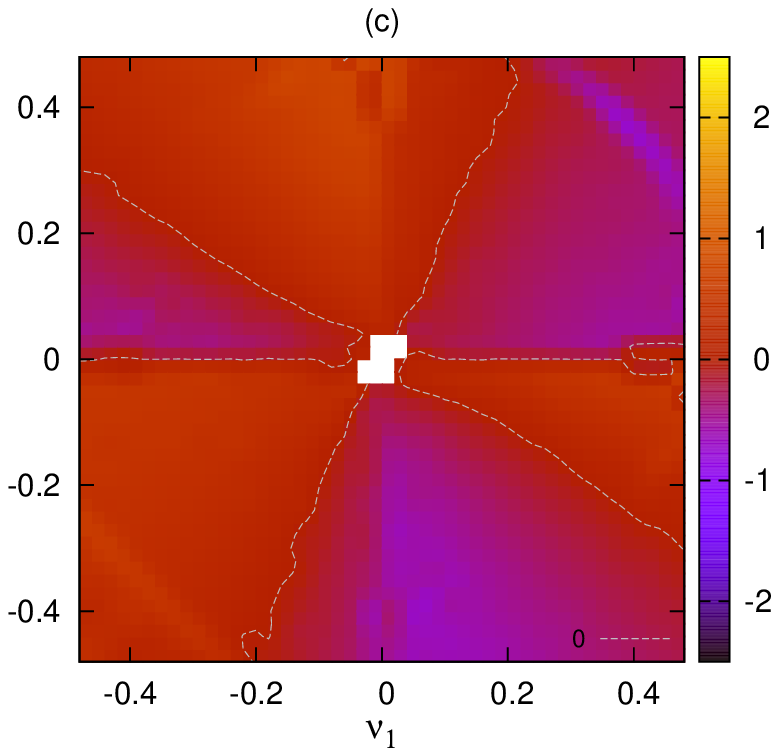}
\end{center}
\caption{\label{comp}
(Color online) Compressibility of electrons in the two graphene layers in the double-layer structure.
The compressibility map for (a) $\frac{\partial\mu_1}{\partial\rho_1}$,
(b) $\frac{\partial\mu_2}{\partial\rho_2}$,
and (c) $\frac{\partial\mu_1}{\partial\rho_2} = \frac{\partial\mu_2}{\partial\rho_1}$, in units of $E_0\ell^2$. }
\end{figure}

\subsection{Electronic compressibility}

One experimental consequence of the formation of the CDW states would be the negative contribution of the e-e correlations to the electronic compressibility in the layers, similarly to the case of monolayers screened by a gate.\cite{shk} This is measurable via the difference 
\begin{equation}
\chi_{ij}=\frac{\partial\mu_i}{\partial\rho_j}=
\left(1+\delta_{ij}\right)\frac{\partial E_{\text{coh}}}{\partial\rho_i}+
(\rho_1+\rho_2)\frac{\partial^2 E_{\text{coh}}}{\partial\rho_j\partial\rho_i},
\end{equation}
between the actual differential capacitance of a device sketched in Fig.~\ref{comp},
\begin{equation}
\frac{\partial V_j}{\partial\rho_j}=(\hat C^{-1})_{ij}+\frac{\chi_{ij}}{e^2},
\end{equation}
and its classical geometric capacitance $(\hat C^{-1})_{ij}$. This results from several contributions to the chemical potential of the double-layer system,
\begin{equation}
\mu_i=\text{const.}+E_{\text{coh}}+(\rho_1+\rho_2)\frac{\partial E_{\text{coh}}}{\partial\rho_i},
\end{equation}
which includes the kinetic energy, the exchange interaction energy among the states in the deeper-lying completely filled LLs,
the exchange interaction of the partially filled $n_1$ and  $n_2$ LLs with the sea of filled LLs,
and the cohesive energy $(N_1+N_2)E_{\text{coh}}$ of the partially filled LLs $n_i$.
Figure \ref{comp} illustrates the contributions of the electrons' compressibility $\chi_{ij}$ to the capacitance in the CDW states, for the same parameters as in Fig.~\ref{phases}.

Another experimental consequence of the formation of CDW states would be the reentrant integer quantum Hall effect behavior of electrons in such system, determined by pinning the hexagonal CDW domains and domain walls by disorder. 

\subsection{Connection to other systems}

So far we have focused on electron-electron or the equivalent hole-hole double layers.
If one layer is $p$ doped while the other is $n$ doped (e-h bilayer), the second term in Eq.~(\ref{coh}) changes sign.
In the latter case, the cohesive energies of stripe and square/rectangular CDWs for e-e and e-h layers are related, but a connection between e-e and e-h double-layer CDW with rhombic/hexagonal lattice is nontrivial;
hence, the study of the phase diagram for the e-h double-layer CDW will be reported in a separtate paper.
Another point to make is related to the electron tunneling between graphene flakes.
In G/hBN/G heterostructures produced by mechanical exfoliation and transfer,
the graphene layers are rotationally misaligned.
Then, the two graphenes' Brillouin zones are rotated with respect to each other, displacing the Dirac points on the momentum plane by $Q\gg\ell^{-1}$, so that interlayer tunneling would be resonant and could affect the electron spectrum only at high energies $\sim vQ$,\cite{2LG} irrelevant for the formation of low-energy Landau levels. As a result, even for the thinnest interlayer separation (with only one hBN layer between graphene flakes), the interlayer tunneling can be neglected in the analysis of the CDW phases in the two-layer system. 

\section{Summary}
\label{summary}

We have shown that the two-dimensional electron gas in G/hBN/G heterostructures in a perpendicular magnetic field
has several correlated charge-density-wave phases, and the critical temperature of such ordering is
in the experimentally accessible temperature range.
These features can be probed in both transport and capacitance measurements.

\subsection*{Acknowledgement}

This research was funded by the European Graphene Flagship, the Hungarian Academy of Sciences,
Hungarian Scientific Research Funds No.\ K105149,
Royal Society Wolfson Research Merit Award,
ERC Advanced Grant ``Graphene and Beyond'',
and ERC Synergy Grant ``Hetero2D''.
Numerical computation was performed using the HPC facilites at the Budapest University of Technology and Economics.

\clearpage
\onecolumngrid

\renewcommand{\thefigure}{S\arabic{figure}}
\renewcommand{\thetable}{S\arabic{table}}
\setcounter{equation}{0}
\setcounter{figure}{0}
\setcounter{section}{0}

\section*{Supplementary Online Material to ``Charge-density waves in double-layer graphene structures in a high magnetic field''}

\section{Analysis of the optimized parameters}

The order parameters of our calculation are the Fourier components of the guiding-center density, c.f.\ Eqs.~(1-2).
We make an Ansatz for the shape of the guiding-center density modulation, namely, that it is a
rectangular wave in the stripe phase and a Bravais lattice of sharply bounded circular disks in the
bubble crystal phases (c.f.\ the top panels of Fig.~1).
The actual charge density is determined by the guiding-center density and the shape of the Landau orbitals in the
partially filled Landau level.
The Ansatz contains up to three parameters to be optimized numerically.

For the stripe phase, these are the period (wave length) $a$ and the relative shift of the
two stripe structures in the two layers, $\bdelta$.
The latter is assumed to be along a direction that is perpendicular to the stripes (c.f.\ the top left panels of Fig.~1).
Notice that the ratio of the filled and unfilled parts of the CDW is determined by the filling factor, and
it can be different in the two layers; the same holds for all CDWs we consider.

For the square and hexagonal CDWs, the parameters are the length of the primitive lattice vectors
$a=|\mathbf a_1|=|\mathbf a_2|$ and the relative shift $\bdelta$.
The latter is now assumed to be along the vector $(\mathbf a_1+\mathbf a_2)/2$ for the square lattice,
and along $(\mathbf a_1+\mathbf a_2)/3$ for the hexagonal lattice.
These vectors point from a vetrex to the midpoint of the nearest square or triangle, respectively
(c.f.\ the top center and right panels of Fig.~1).

In our calculation we also considered two more two-dimensional Bravais lattices, the rectangular
and the rhombic (centered rectangular), although as a result we found that these do not occupy any phase volume.
Here we optimized three parameters: apart from $a$ and $\bdelta$, the length ratio of the
primitive lattice vectors, $\beta=|\mathbf a_2|/|\mathbf a_1|$ was used.
(There are, of course, other equivalent parametrizations using angles.)
$\bdelta$ was assumed to be along the vector $(\mathbf a_1+\mathbf a_2)/2$ for the rectangular lattice.
For the rhombic lattice, this direction was along the vector from a vertex to a nearby point that has an equal
distance from all of the three nearest vertices, $\frac{a}{2}\mathbf{\hat x}+
\frac{a}{4}\left(\sqrt{4\beta^2-1}-\frac{1}{\sqrt{4\beta^2-1}}\right)\mathbf{\hat y}$.

\subsection{The shift between the charge-density wave in the two layers $\bdelta$}

\begin{figure}[htbp]
\begin{center}
\includegraphics[width=0.32\columnwidth,keepaspectratio]{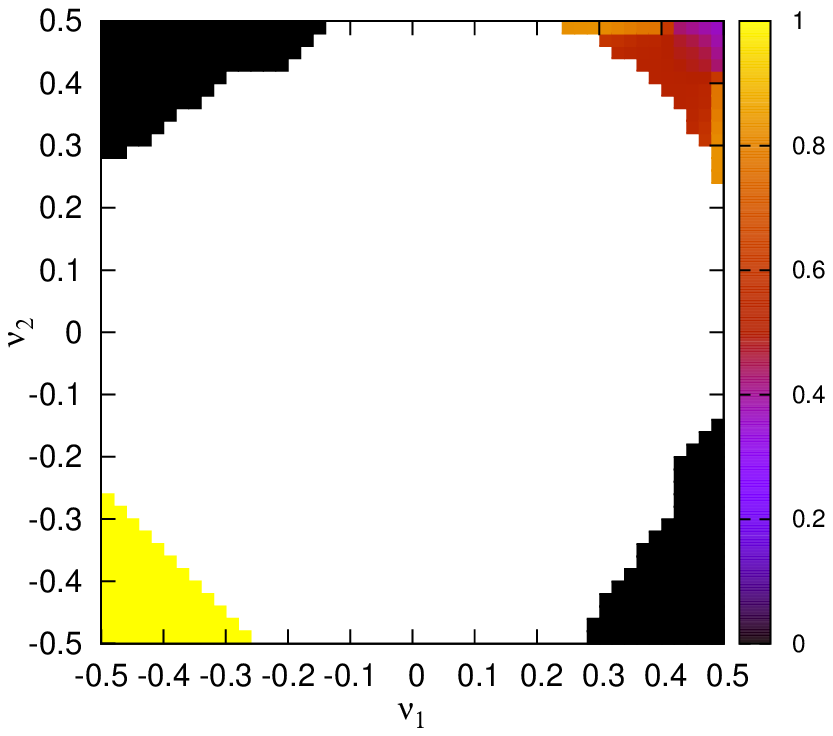}
\includegraphics[width=0.32\columnwidth,keepaspectratio]{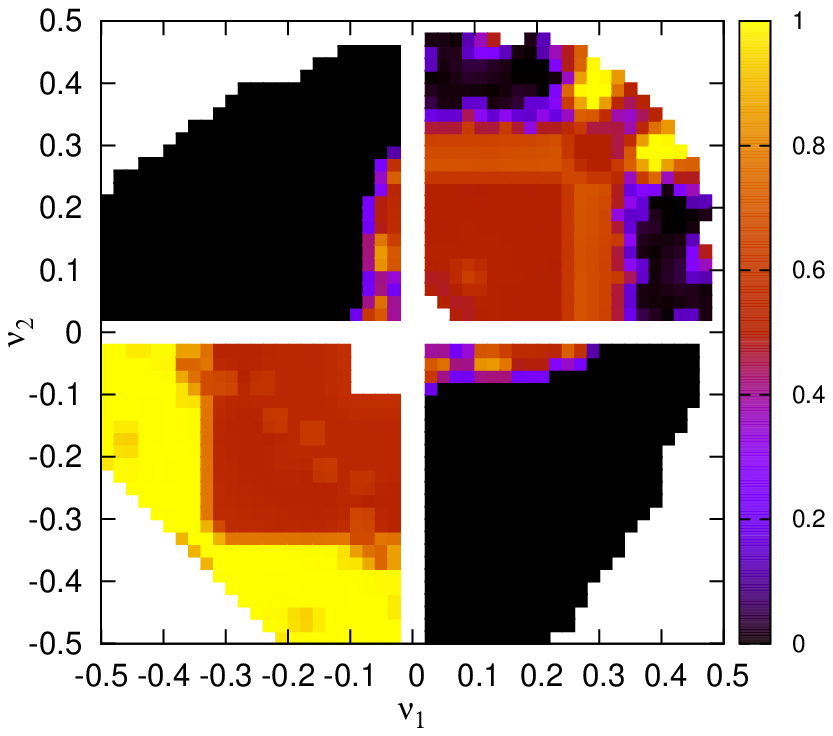}
\includegraphics[width=0.32\columnwidth,keepaspectratio]{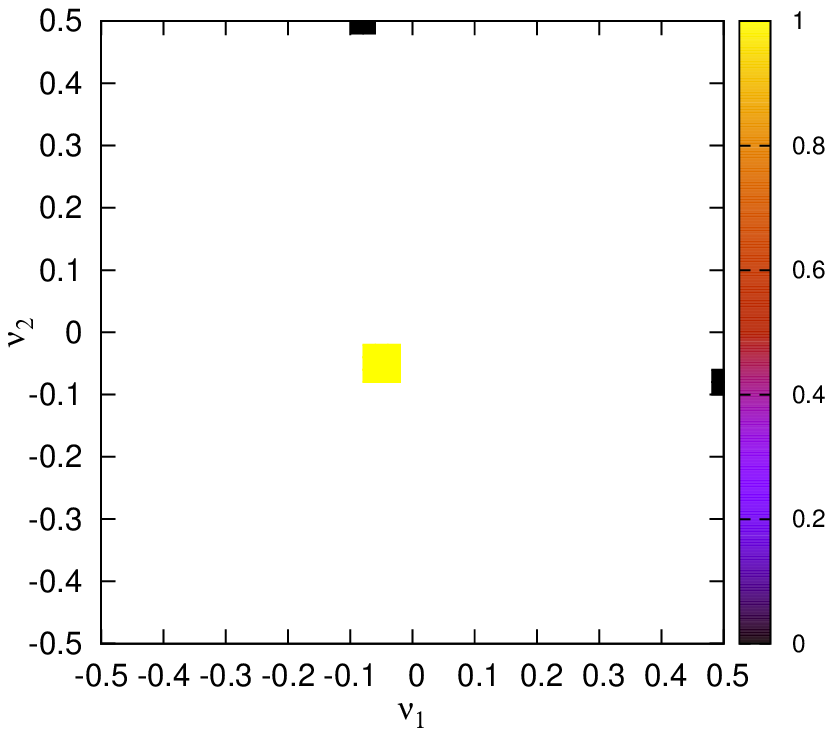}
\end{center}
\caption{\label{delta}
The shift between the charge-density wave in the two layers $\bdelta$
(a) in the stripe phase, (b) in the rhombic/hexagonal phases, and (c) in the rectangular/square phases.
The points where $\overline\nu_1=0$ or $\overline\nu_2=0$ have been removed, as  $\bdelta$ is meaningless there.
}
\end{figure}

In the``excitonic region'', i.e., if $\overline\nu_1$ and $\overline\nu_2$ have opposite sign,
the charge-density modulations in the two layers attract each other.
We find $\bdelta=0$ in the stripe [Fig.~\ref{delta}(a)] and the square [Fig.~\ref{delta}(c)] lattice phases.
In the hexagonal phase [Fig.~\ref{delta}(b)], $\bdelta=0$ almost everywhere,
except for a small region where it is positive but small (small hole-doping in one layer and moderate
electron doping in the other).
This feature must be connected to the complex shape of charge densities at low doping.

If $\overline\nu_1$ and $\overline\nu_2$ are both negative, $|\bdelta|=a/2$ for the stripe and
$\bdelta=(\mathbf a_1+\mathbf a_2)/2$ for the square lattice;
for the hexagonal phase $\bdelta=(\mathbf a_1+\mathbf a_2)/3$ for large doping
but decreases somewhat where both $|\overline\nu_1|$ and $|\overline\nu_2|$ are small.
For the stripe and square lattices this is intuitive, as this shift corresponds to the maximal distance of the
regions of high guiding-center density in the two layers.
The shortening of $\bdelta$ in the hexagonal CDW at small doping is less obviously intuitive,
but the Coulomb repulsion is between charge-densities and not guiding centers.
In Fig.~2(b) of the letter we see a case where the nontrivial charge-density profile stabilizes a
$|\bdelta|\neq a/2$ shift for the stripe; the optimal shift allows the highest charge-density bump in one layer
to coincide with the deepest dip in the other layer.

If $\overline\nu_1$ and $\overline\nu_2$ are both positive, $\bdelta$ is nontrivial (neither zero nor maximal)
both in the hexagonal and the stripe phases [Fig.~\ref{delta}(a,b)].
Notice that the $n=2$ Landau orbitals, which connect the guiding-center density to the charge density in
this region, are rather complex.
Actually, Fig.~2(b) was taken from this part of the phase space.

\subsection{The period or wave length $a$}

\begin{figure}[htbp]
\begin{center}
\includegraphics[width=0.32\columnwidth,keepaspectratio]{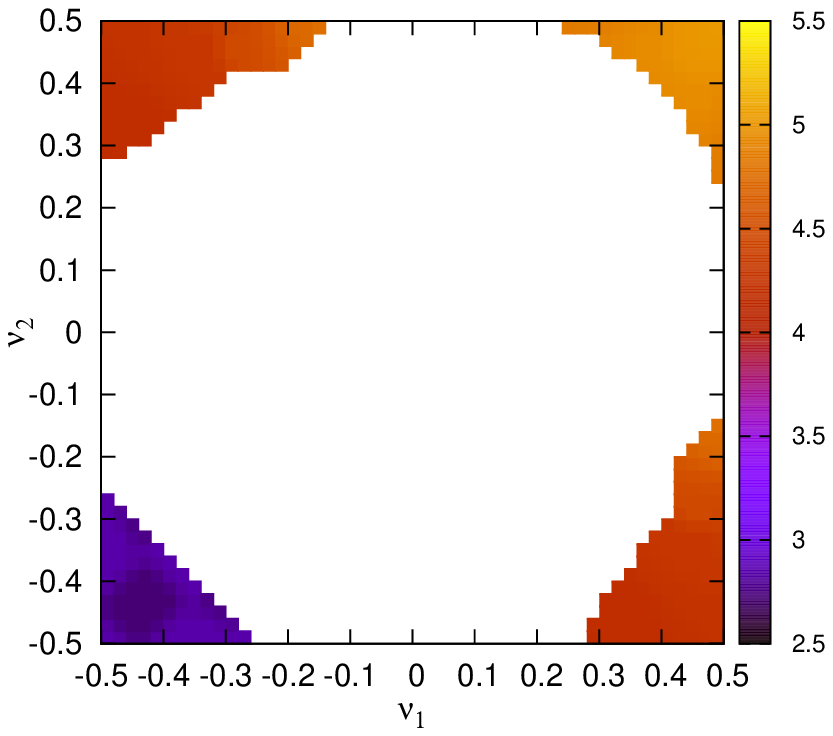}
\includegraphics[width=0.32\columnwidth,keepaspectratio]{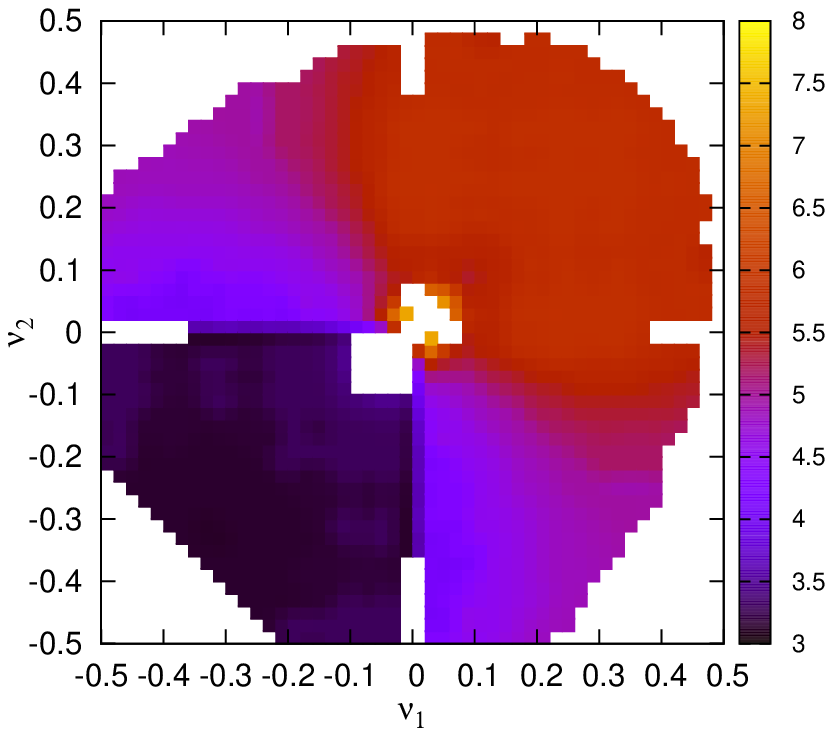}
\includegraphics[width=0.32\columnwidth,keepaspectratio]{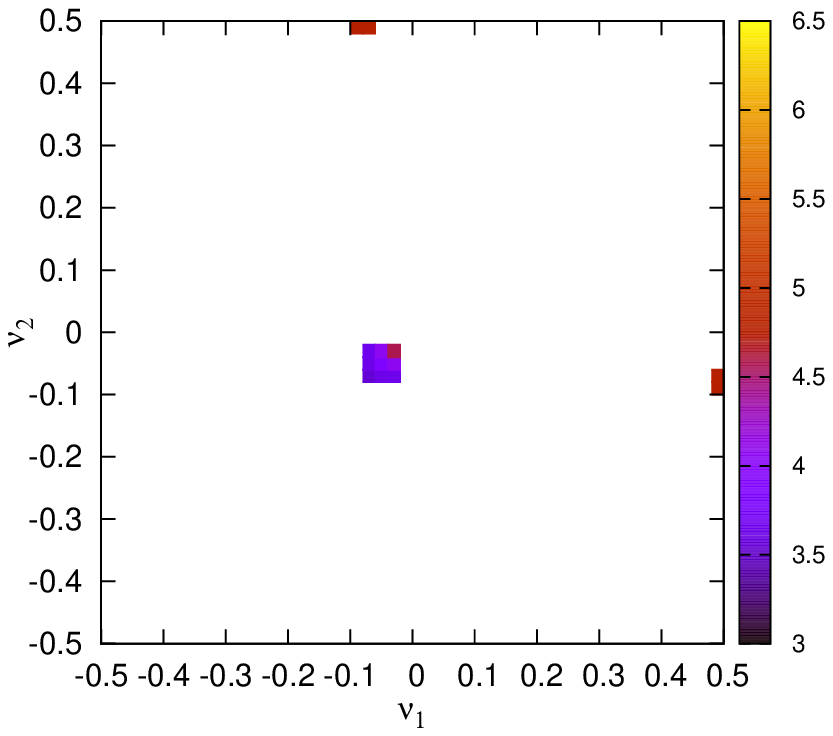}
\end{center}
\caption{\label{lambda}
The period (wave length) of the charge-density wave $a=|\mathbf a_1|$
(a) in the stripe phase, (b) in the rhombic/hexagonal phases, and (c) in the rectangular/square phases.
In the rectangular phase $a$ is the shorter side of the rectangular unit cell.
The rhombic lattice can be regarded as center rectangular; $a$ is the shorter side of this rectangle.
The length unit is the magnetic length $\ell$.
}
\end{figure}

Fig.~\ref{lambda} shows the period (wave length) $a$ in the phases we consider, in units of the magnetic length $\ell$.
In the rectangular phase we define $a$ as the length of the shorter primitive lattice vector.
The rhombic lattice can also be regarded as center rectangular; we define $a$ as the shorter side of this rectangle.
The cyclotron radius is $R_c=\sqrt2\ell$ in the $n=1$ Landau level and $R_c=2\ell$ in the $n=2$ Landau level.

The periods we find are comparable to those in single layer systems, where $a/R_c$ is typically between 2.3 and 2.8.
In the excitonic region, where the cyclotron radii in the two layers differ, the wave length of CDWs
are between those of the corresponding phases in the $n=1$ and $n=2$ Landau level regions.

Where the hexagonal/rhombic CDW connects to the square CDWs that probably precursor the Wigner crystal,
i.e., at small positive doping, we observe an elongation of the period of the hexagonal phase.
This, together with the rhombic deformation (see Subsec.~\ref{secbeta} below), suggest that the hexagonal
and the square CDWs in the $n=2$ quadrant ($0<\overline\nu_1,\overline\nu_2$) are connected by a second-order
phase transition.

\subsection{The ratio of the two primitive vectors $\beta=|\mathbf a_2|/|\mathbf a_1|$}

\label{secbeta}

The optimized value of $\beta$ is always unity within numerical error in the rectangular phase,
which simply means the square CDW is stable.
In most of the rhombic phase we found the same, although the numerical noise was slightly higher.
There is one exception: near the arc that is occupied by the square CDW in the 
$n=2$ quadrant ($0<\overline\nu_1,\overline\nu_2$) $\beta$ of the rhombic phase smoothly
decreases from 1 to $1/\sqrt{2}$, suggesting a continuous quantum phase transtion.

\begin{figure}[htbp]
\begin{center}
\includegraphics[width=0.32\columnwidth,keepaspectratio]{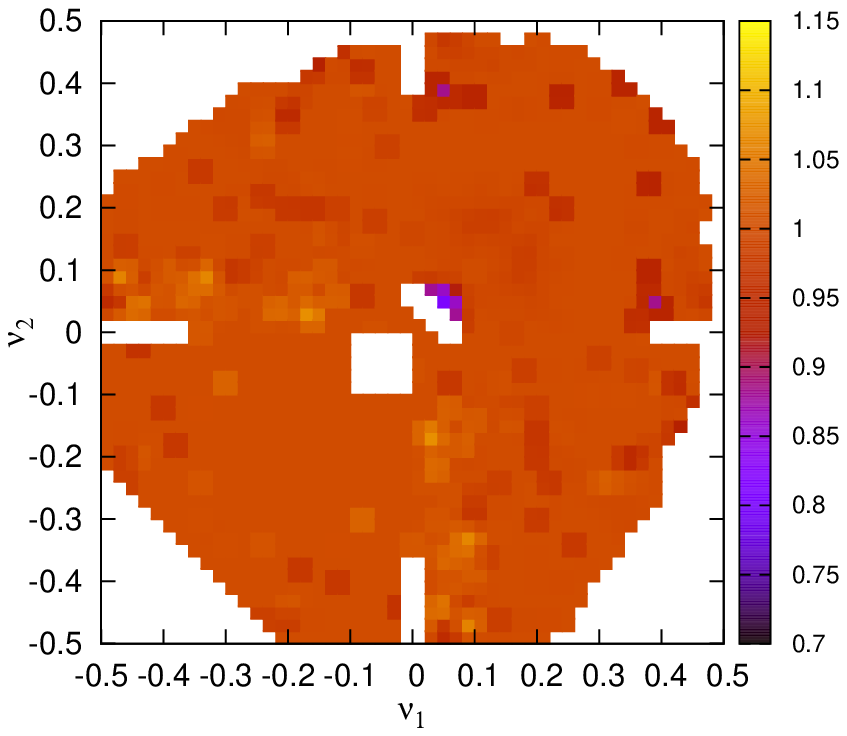}
\includegraphics[width=0.32\columnwidth,keepaspectratio]{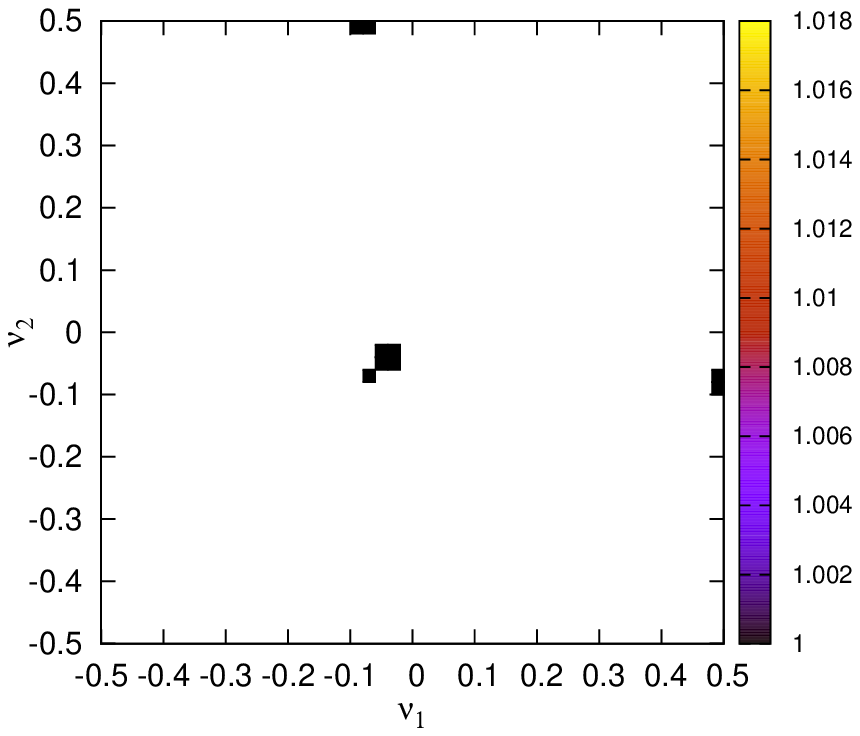}
\end{center}
\caption{\label{beta}
The ratio of the two primitive vectors $\beta=|\mathbf a_2|/|\mathbf a_1|$
(a) in the rhombic/hexagonal phases and (b) in the rectangular/square phases.
The special value $\beta=1$ corresponds to the hexagonal lattice in the first case, and
to the square lattice in the second case.
}
\end{figure}


\begin{thebibliography}{99}
\bibitem{footnoteE} 
Note that $E_0 = 111\sqrt{B\text{ [T]}}$ K, so that $E_0 \sim 300$K at $B \sim 10$T. Here we use  $\epsilon_\parallel=5.05$ and $\epsilon_\perp=6.85$ for hBN.
\end{thebibliography}
\end{document}